# Fundamentals and applications of aberration corrected high resolution transmission electron microscopy in materials science

Ranjan Datta[1,*], Sneha Kobri M.[1], and Sudip Mahato[1]

[1]*International Center for Materials Science, Jawaharlal Nehru Centre for Advanced Scientific Research (JNCASR), Jakkur, Bangalore 560064, India.*

**Abstract**

In this review article fundamentals of aberration corrected phase contrast transmission electron microscopy for the structural characterization of materials at atomic length scale is presented. The word 'structure' entails atomic arrangement as well as electronic structure information of the materials. The article summarily covers a range of topics on the basics of aberrations, aberration correctors, direct image interpretation with negative $C_S$ phase contrast microscopy, a discussion in comparison with the competitive atomic resolution phase contrast methods e.g., off-axis electron holography, electron ptychography, differential phase contrast microscopy. Additionally, various examples of quantitative imaging of materials at atomic length scale, associated image simulation and reconstruction methods for retrieving the phase information are presented. With the tremendous advancement in instrumentation and recording devices, potential future perspective of such tools and methods in solving challenging materials science problems are outlined.

[*]Corresponding author's email: ranjan@jncasr.ac.in

**Contents**

**I. Introduction**

In this review article, an outline is provided based on the author's presentations in several conferences and meetings held in Electron Microscopy Society of India (EMSI 2024, 2025), Indian Institute of Technology Madras (IITM 2017, 2022, 2025), JNCASR (2022) and related publications on the topic [1–3]. It is a blend of conventional subjects and author's own contribution and experience from working in the field of transmission electron microscopy for more than 20 years. The review will be self-contained, and references are cited in the text to guide the readers on already published illustrations.

High resolution transmission electron microscopy (HRTEM) or atomic resolution phase contrast imaging technique enables recording of periodic arrangement of atoms in a crystal along high symmetry orientations [Fig.1]. It is possible to intuitively infer on the crystalline quality, basic symmetrical properties, interplanar spacings and even identification of various crystallographic defects in materials from such images [4–6]. Some typical examples of HRTEM images recorded under negative $C_S$ imaging condition are: a) NiFe$_2$O$_4$ along [001] zone axis (Z.A.) directly showing the spinel cubic symmetry and identification of areas with A site cation vacancy associated with dark diffused contrast areas [7–11], b) cross sectional view of MoS$_2$ van der Waals epitaxial thin film grown on *c*-plane sapphire showing 2H-polytype stacking, orientation relationship between the film and the substrate (<11-20>Al$_2$O$_3$ || <01-10>MoS$_2$) and interfacial atomic bonding between Al and S atoms [12], c) MoS$_2$-ReS$_2$ 2D-alloy showing phase transitions from 2H to various forms of 1T$_d$ depending on the Re concentration [13,14]. In section-II, various experimental and computational methods are presented which are essential to extract intricate information from such images e.g., three-dimensional arrangement of atoms in nanocrystals [15], chemical identification of atoms [16,17], chemical bonding state of a dopant in the host lattice [18] etc. It is relevant to mention that HRTEM has an equivalent technique using scanning mode of electron beam popularly known as Z-contrast or STEM-HAADF imaging to obtain atomic resolution images. Image formation in Z-contrast imaging is incoherent whereas in HRTEM it is coherent. One extreme capability of Z-contrast imaging is demonstrated in the case of direct imaging of light atoms C in graphene lattice with brighter dots in the image is identified as Si and N dopants [19]. Additionally, STEM-HAADF technique has the advantage of obtaining simultaneous electron energy loss (EEL) spectra from the respective atomic columns enabling chemical identification, bonding state and electronic structure information. The near edge structure of EEL spectra is shown to be markedly different depending on the distinct dopant location. However, as already mentioned above, the HRTEM method is approaching similar capabilities entirely through the imaging route and presently the reports are scarce in this area.

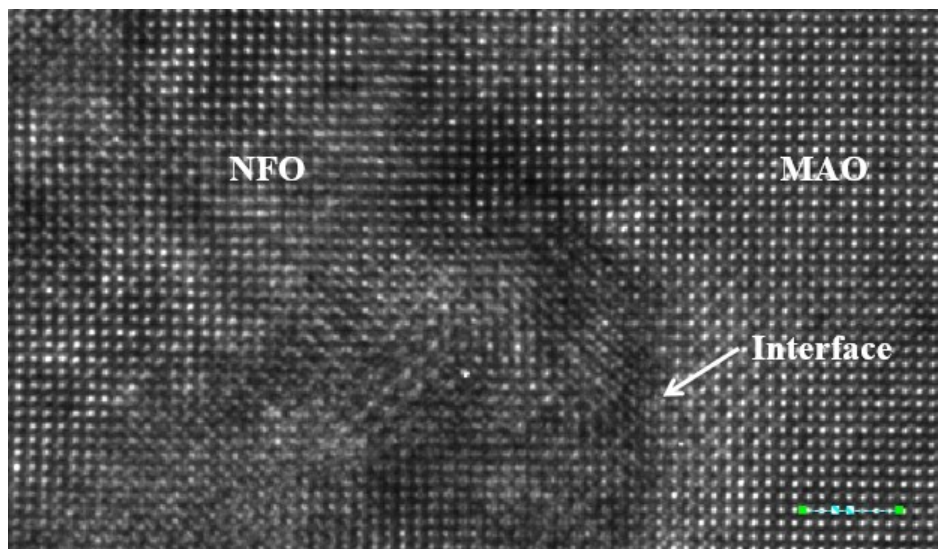

**Fig. 1.** *Example HRTEM image of NiFe$_2$O$_4$ thin film grown on MgAl$_2$O$_4$ substrate showing formation of coherent interface between the two crystal lattices. Distance between two bright dots in MAO is ~2.17 Å.*

### A.  Phase contrast microscopy

The phase contrast microscopy was originally discovered by F. Zernike in an optical microscope using a phase plate made of thin metallic foil. Zernike was awarded Nobel prize in physics in 1953 for the discovery. He could enhance the contrast of biological tissue samples significantly without any chemical staining [20]. There is a marked difference between the bright field (BF) image and the phase contrast image from a weakly scattering tissue sample. Without chemical staining, specimen and background appear almost with the same brightness in a BF image but improve significantly in the phase contrast image with more details. One can understand the principle of phase contrast with the help of the phasor diagram following the description by Zernike. In Fig. 2 (a), OR and OR´ represent the waves or vibrations, where *x*-component of the vibration gives the amplitude and angle from the *x*-axis gives the phase. Now for a strong scatterer, the incident wave OA changes to OS [Fig. 2(b)] and the compounding effect is OS´. Thus, scattering regions will appear dark in a white background in a BF image. This is the amplitude contrast. Now for the weak scatterers, or for a weak phase object, the incident wave changes from OA to OS [Fig. 2(c)]. There is no change in amplitude except the phase of the illumination, and the compounding effect is OS´. The effect will not be visible unless it is translated into an amplitude contrast. This is performed with the help of a phase plate by rotating the phase of either OA or OS to OU´ and the compounding effect between the OU´ and OS´ will be translated into dark contrast in the image.

However, Zernike like phase plate as described above is not available for the electron microscope. Instead, combination of both $C_s$ and $\Delta f$ are used in the form of phase contrast transfer function (PCTF) function to introduce phase contrast during HRTEM imaging [Fig. 3]. Moreover, few other distinct atomic resolution phase contrast methods are available e.g., off-axis electron holography, DPC, iDPC [21–25] etc. In this article, primarily HRTEM phase contrast technique is focused with a discussion later in comparison with off-axis electron holography followed by outlining fundamental principles of DPC, iDPC, and ptychography imaging techniques for the general interest of the readers.

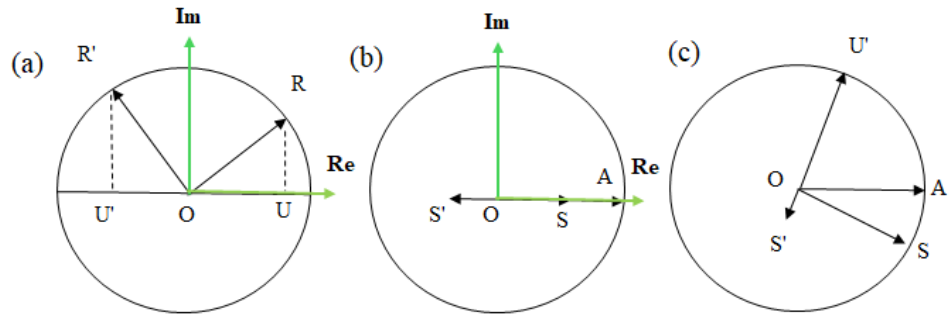

**Fig. 2.** *Understanding the origin of phase contrast with the help of phasor diagram following Zernike's description.*

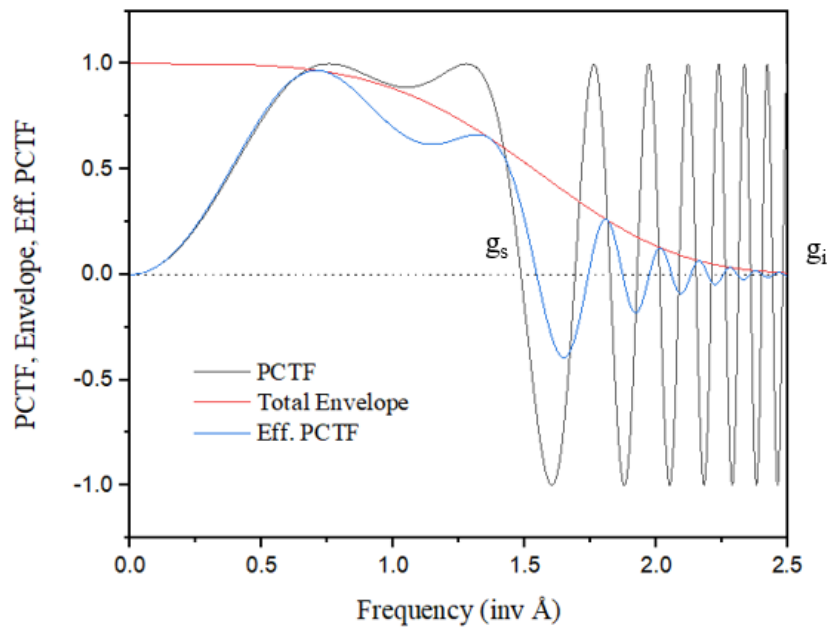

**Fig. 3.** *PCTF function as phase plate for HRTEM. Black line is the PCTF without envelope, red line is the example combined envelope function, and the blue line is the PCTF after considering envelope function.*

**B.    Definition of spatial resolution**

The definition of spatial resolution of an ideal lens is presented before proceeding for the description of aberration and its effect on spatial resolution in the context of HRTEM. In general, spatial resolution is defined by a minimum separation distance $d_0$ by which one is just able to distinguish or discern between two points or lines [Fig. 4 (a)]. Unaided human eyes can resolve maximum 100-200 µm and require an optical microscope to see features smaller than that. Due to wave nature of the illumination and consequently the self-interference-based diffraction effect, a point object is always blurred into a disk known as Airy pattern [Fig. 4 (b)]. The radius of the central peak of the airy pattern is given by the Eq. 1 and is governed by the wavelength of the illumination ($\lambda$) and the semi-angle ($\beta$) of the collection aperture. Now, Rayleigh's criterion for spatial resolution is defined in terms of minimum discernible distance between two such airy disks, i.e. by $d_0$, where the first minimum of one peak fall right below the maximum of the second peak [Fig. 4 (c)]. However, this is not a strict criterion, and one can also use situations where two peaks can slightly be merged to extract more information. Such a situation has already been used in the context of HAADF imaging. For an optical microscope the $d_0$ is given by the Eq. 2 considering numerical aperture for both the condenser and objective apertures.

$$\delta_D = 0.61 \frac{\lambda}{\alpha} = \frac{0.61\,\lambda}{\mu \sin \beta} \tag{1}$$

$$d_0 = \frac{1.22\,\lambda}{NA_{Obj} + NA_{Con}} \tag{2}$$

The maximum details that can be observed by using an optical microscope depend on the wavelength ($\lambda$) of the light and the numerical aperture ($\mu \sin \beta$). This is the limit of the information given, no other blurring due to aberration and other disturbances are present. Shorter the wavelength and larger the aperture opening, smaller will be the radius of Airy disk or smaller the diffraction broadening and better the resolution. In optical microscope, lenses can be made to near perfection. Thus, the performance of a diffraction limited visible light optical microscope can further be improved in an electron microscope where much shorter wavelength of electron as illumination can be used. Louis De Broglie showed that the wavelength of electron can be tuned with the accelerating potential i.e. $\lambda \approx 1.22 \times E^{-0.5}$, thus the $\delta_D$ can be improved accordingly. For example, at 100 kV applied anode potential, the wavelength of electron is 4 pm, which is one order of magnitude smaller than the size of the atom. However, for practical situations decreasing the wavelength of electron involves increasing its kinetic energy and limit is posed by the tolerable sample damage. On the other hand, increasing the aperture size increases the aberration in the electron lens. Moreover, given a finite size requirement of the instrument limits the size of the numerical aperture. Therefore, there is an optimum tradeoff required and consequently correction of aberration of electron lenses becomes important.

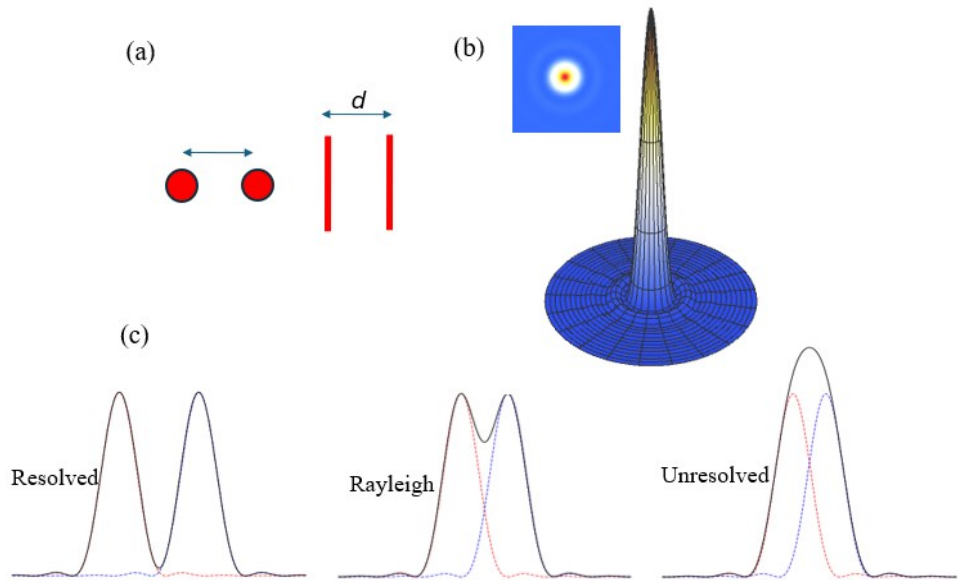

**Fig. 4.** *(a) Distance between two distinct spots and lines, (b) Airy pattern, and (c) Rayleigh's criterion.*

Now, in the context of HRTEM, spatial resolution is limited by the lens aberrations and is defined by the phase contrast transfer function (PCTF). An example PCTF [$C_s = -44 \ \mu m, \Delta f = 6 \ nm$] is shown in Fig. 3 (a). The ideal shape of PCTF should be to minimize the gradient for a given frequency band width to reduce the delocalization of the image. The first zero crossing of PCTF $g_S$ on the x-axis defines the point resolution and the $g_I$ is the information limit. However incoherent aberrations, such as chromatic or temporal aberrations, spatial coherence of the source, instrumental instability, and environmental noise appear as incoherent envelopes and limit the information limit of the microscope. If these are significantly high, then they can limit the point resolution of the microscope [Fig. 3 (c)]. Incoherent aberrations were the resolution limiting factor during Scherzer's time, when his student Robert Seelinger built the first aberration corrected microscope [26]. Their work involved demonstrating the working principle of aberration correctors on reducing the effect of $C_s$ rather improving the actual resolution of the microscope. In an aberration corrected electron microscope, $C_s$ can be set to 0, but there will be always some resolution limiting higher order aberration present. A small amount of $C_s$ is also required for setting up a suitable PCTF for a meaningful HRTEM imaging experiment.

### C. Aberrations

Aberrations is the blurring of the focal point of the rays originating from the same object point [Fig. 5]. Thus, a point in object is blurred in the image plane. Blurring may be caused by several factors: geometrical imperfections of lenses and this in light optics are equivalent to non-uniform lens action across the plane of the magnetic pole piece, chromatic aberration due to spread in the energy in the source illumination ($\Delta\lambda$), environmental instabilities originating at mechanical vibrations, electromagnetic stray fields, temperature fluctuations, acoustic sources etc. Geometrical aberrations are coherent aberrations and do not cause permanent loss of information and the diffused information can be recovered through appropriate image reconstruction techniques. On the other hand, the information is lost permanently due to incoherent aberrations. Mechanical vibration, finite size of electron source, instability in power supply, specimen drift, and point spread function of detectors are parasitic aberrations and they can be either coherent or incoherent. For example, two-fold astigmatism is a coherent parasitic aberration; mechanical vibration is an incoherent parasitic aberration. Geometrical aberration can be corrected by aberration correctors. Chromatic aberration can be improved with the monochromator or using a Cc corrector system. Instrumental instability can be improved by improving the respective components.

Given the environmental requirement for the microscope installation and other criteria have been met to the desired specifications for an ultra-high resolution transmission electron microscope, the spatial resolution will then be governed by the limiting geometrical aberration of the lens. The effect of two most encountered aberrations of electromagnetic lens, spherical and chromatic aberrations are shown in Fig.5. In case of spherical and chromatic aberrations, blurring is caused due to the inhomogeneous magnetic field and the finite energy spread of the electron source, respectively and results in decrease in point resolution. Slight adjustment in defocus can be used to reduce the effect of blurring due to spherical aberration by projecting the disk of least confusion to the Gaussian image plane [Fig. 5 (a)]. This is performed by Scherzer's defocus condition in an uncorrected microscope with the optimized phase contrast transfer function (PCTF). The effect of both the aberrations are the same on broad beam HRTEM and scanning probe mode STEM.

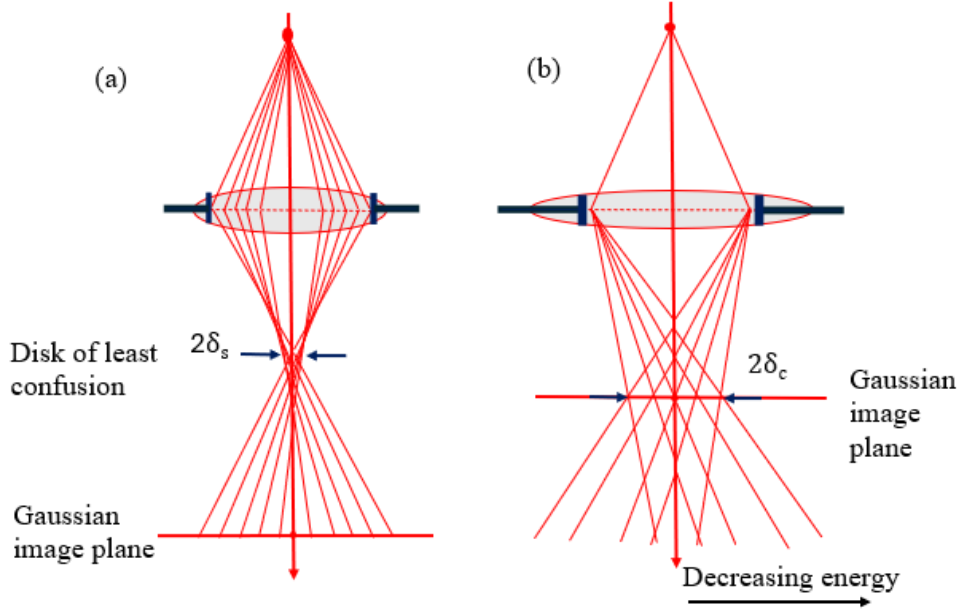

**Fig. 5.** *Two most encountered aberrations in transmission electron microscope: (a) spherical $\delta_S = \frac{1}{4}C_3\alpha^3$ and (b) chromatic $\delta_C = C_C\alpha\frac{\Delta E}{E_0}$ aberrations.*

The definition of third order spherical aberration ($C_3$) for a round symmetrical lens is given in Fig. 6. Under paraxial approximation all the rays emanating from $P_0$ at small angle $\gamma$ will meet at point $P_1$ on the optic axis. $L_0$ is the focal length of the lens. Paraxial means electron follows flat trajectories near the optic axis i.e. $\cos\gamma \approx 1$. Within this small angle approximation all the rays will be focused at $P_1$. Aberration appears due to non-ideal imaging systems beyond paraxial approximation which can be described mathematically by expanding $\gamma$ to the higher order terms following Taylor series expansion and up to second order is given by $\cos\gamma \approx 1 - \frac{\gamma^2}{2}$. Under this condition, the rays will not be focused on the Gaussian image point but will be displaced at a distance $r_s$. For spherical aberration one can compute $L_z = \frac{2\pi m_0 v}{eB_z}(1 - \frac{\gamma^2}{2})$ and longitudinal spherical aberration $\Delta s = L_z - L_0 = -\frac{1}{2}L_0\gamma^2$. With the help of geometry, the relationship between $r_s$, $L_0$, and $\gamma$ can be derived and is given by $r_s = \Delta s \tan\gamma \approx \Delta s\gamma = -\frac{1}{2}L_0\gamma^3$. Finally, third order spherical aberration coefficient is given by $C_3 = -\frac{1}{2}L_0$ which depends on the fixed focal length of the lens and thus is the universal constant of the lens. It is called third order because it depends on the third power of the inclination angle. However, there are many other types of geometrical aberrations of various orders and require a generalized approach to characterize them.

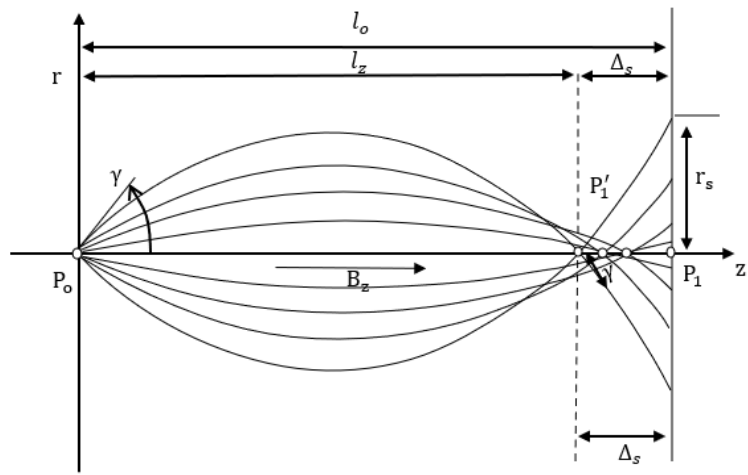

**Fig. 6.** *Third order spherical aberration from the perspective of deviation from paraxial ray path.*

The definition of the image aberration is given in Fig. 7. Paraxial rays require two parameters to define the trajectories; these are the positions of rays at the object ($w_o$) and the aperture planes ($\omega$). Knowing these two parameters, it is possible to find the position $w_i$ of the ray in the image plane which is the reflection of $w_o$. However, for aberrated ray the position of the ray will be at $w_i'$, and $\Delta w_i$ defines the image aberration. Image aberration is the difference in distance between the Gaussian image position to the actual position of the ray. Each aberration type has its own characteristic aberration figure which may depend on the object position $w_o$ and scattering angle $\omega$. There are aberrations which depend only on the scattering angles, these are called axial aberrations. Example of axial aberrations are defocus ($C_1$), $C_3$, chromatic aberration ($C_c$). And the aberrations which depend on the object positions ($w_0$) are called off-axial aberrations.

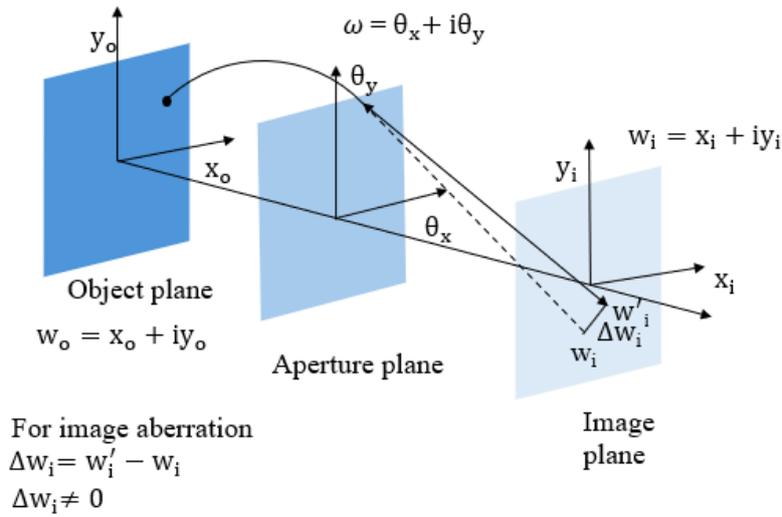

**Fig. 7.** *Definition of image aberration. A ray can be defined by two parameters i.e. scattering angle ω at the scattering plane or w at the image plane.*

There are infinite numbers of geometrical aberrations possible for a round magnetic lens. Number of observable aberration coefficients increases with the precision of the lens and the resolution. In 1856, Ludwig Siedel first analyzed the symmetry permitting aberrations of a round optical lens up to third order. These are known as Siedel aberrations. Due to magnetic field, electron optical systems possess more aberration coefficient in comparison to light optical systems. For optical system and up to third order, the number of aberrations is 5, and in the presence of magnetic field, the number increases to 8. The five aberrations for optical system can be identified as: spherical, coma, field astigmatism, field curvature and distortions. Next, we will go through briefly on these 5 types of aberration for a round magnetic lens and then only listing the allowed aberrations for a non-round hexapole lens-based corrector system which was the one implemented in early aberration corrected microscopes.

Spherical aberration is the first in the list of Siedel aberrations. The aberration figure associated with the spherical aberration is a disk and is centered at the Gaussian image point [Fig. 8(a)]. It is an axial aberration; thus, the aberration figure changes only with scattering angle and remains the same for all the object points. Blurring due to $C_3$ can be reduced by at least ×4 by applying appropriate value of defocus [Fig. 8(b)]. Difference in focal distance between the Gaussian image plane and disk of least confusion is given by $C_{1\,DLC} = -\frac{3}{4} M_L C_3 \omega^2$, where $M_L$ is the longitudinal magnification. The Scherzer defocus for coherent imaging and the optimum value of defocus for incoherent imaging are given by $C_{1\,Scherzer} = -\sqrt{\frac{4}{3}\lambda C_3}$ and $C_{1\,opt} = -\sqrt{\lambda C_3}$, respectively. If one compares the optimum Scherzer focus with $C_{1\,DLC}$ that implies that adjusting the defocus in such a way that the disk of least confusion is moved towards the Gaussian image plane, thus impact of $C_3$ on the imaging process is minimized. This is an alternative way to interpret Scherzer defocus for coherent imaging condition.

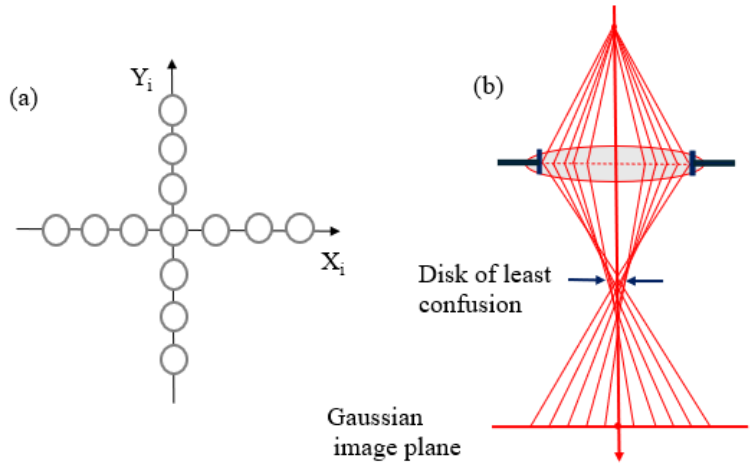

**Fig. 8.** *(a) Spherical aberration is an axial aberration, and blurring is independent of object point, (b) The distance between the disk of least confusion and the Gaussian image place is the amount of defocus required to minimize aberration blurring to large extent.*

Coma is the second in the symmetry permitted geometrical aberration of a round lens and is an off-axial aberration. As the name suggests, the aberration figure resembles like a comet [Fig. 9]. The center of the circle caused by the coma for the ray passes at an angle ω at the aperture plane will be displaced by *l* from the center of the optic axis which is twice the radius of the circle. The *r* and *l* are given by $r = |MB_{31}w_0|\omega\bar{\omega}$ and $l = 2|MB_{31}w_0|\omega\bar{\omega}$, respectively and the combined effect on the image aberration is $\Delta w_i = -M[2B_{31}w_0\omega\bar{\omega} + \bar{B}_{31}\bar{w}_0\omega^2]$. The circles lie on a line and the perimeters of the circle forms 60° angle. The coma is represented by $B_{31}$ and $\bar{B}_{31}$ for radial and azimuthal components, respectively. The first digit indicates order 3 as sum of ray parameters is 3, and the second digit indicates that coma increases linearly with $w_0$. At the center of the optic axis coma vanishes and increases linearly away from the optic axis as indicated by dashed lines. Each object point $w_0$ is imaged into a coma figure. In optical and electrostatic lens, coma will behave similarly. But because of magnetic lens, electron undergoes Larmor precession, coma in magnetic lens has two contributions, radial and azimuthal components. Therefore, two coefficients are necessary to describe coma. Writing in complex notation helps in describing two different components of coefficient by one quantity.

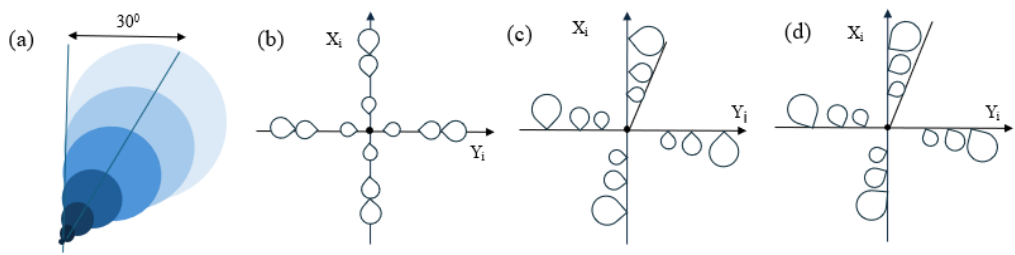

**Fig. 9.** *(a) & (b) Coma, is an off-axial aberration, depends on the object position. (c) Coma aberration figure with azimuthal angle due to magnetic lens. (d) Combined coma aberration figure.*

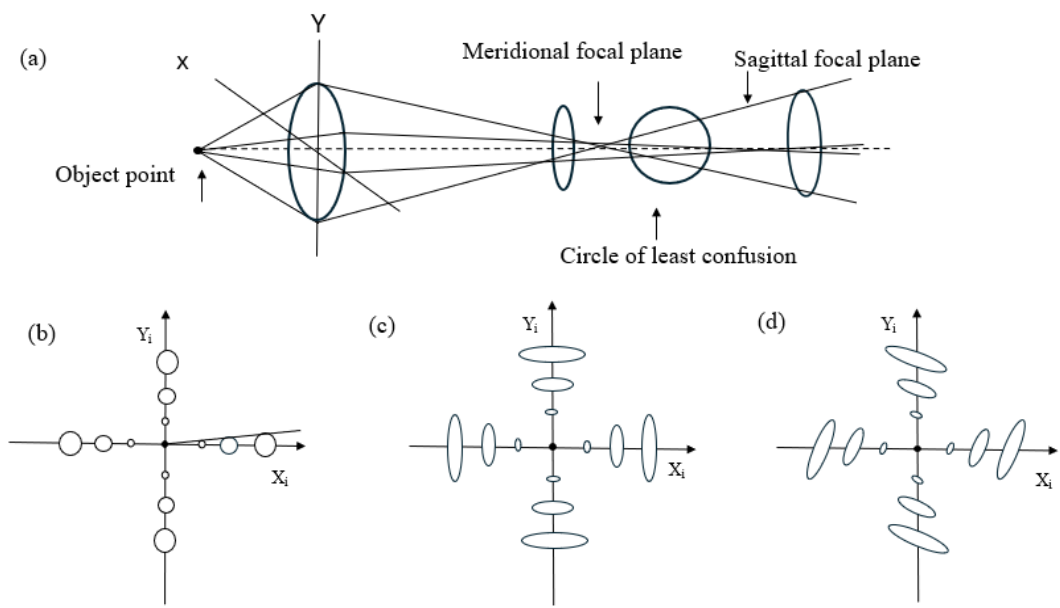

**Fig. 10.** *(a) Two-fold astigmatism. (b) Field curvature, (c) radial components of field curvature and field astigmatism, and (d) combined total effect of field curvature and field astigmatism.*

Field astigmatism ($A_{32}$) is the symmetry permitted aberration of round lens whereas commonly encountered two-fold astigmatism ($A_1$) is a parasitic aberration. Moreover, $A_1$ is an axial aberration and $A_{32}$ is an off-axial aberration. The effect is the same for a given object point. Fig. 10(a) shows the image formation due to two-fold astigmatism. The important point about astigmatism is that the focusing power of the lens depends on the azimuthal angle and is different along orthogonal directions. There are two-line focus planes perpendicular to each other, meridional and sagittal planes. Gaussian image plane lies in between these two planes which is a circle and not a stigmatic image point. Field astigmatism is an off-axial aberration. Due to Larmor precession $A_{32}$ has two component, radial and azimuthal. Aberration figure $\Delta w_i = -M A_{32} \bar{\omega} w_0^2$, changes quadratically from the optic axis. In addition to this, field curvature is another third order aberration which causes the defocus to vary across the field of view with $\Delta w_i = -M F_{32} \omega w_0 \bar{w}_0$. Image plane bends from the optic axis, Field curvature does not have imaginary part unlike field astigmatism. Along the dotted line both vanishes [Fig. 10(b)].

The last in the list of Siedel's aberrations is the image distortions. Image distortion does not cause astigmatism, only the field of view appears distorted unlike the aberrations as described above. Each point finds a stigmatic image point in the image [Fig. 11]. It is a third order aberration. Radial or isotropic image distortion appears either as pincushion or barrel type. The effect of magnetic lens and Larmor precession, azimuthal components to cause the spiral distortion. Radial part is given by real of $D_{33}$ and the azimuthal part is given by imaginary component of $D_{33}$. The aberration figure is $\Delta w_i = -M D_{33} w_0^2 \bar{w}_0$. The overall effect of objective lens on distortion is less but not so by the projector lens.

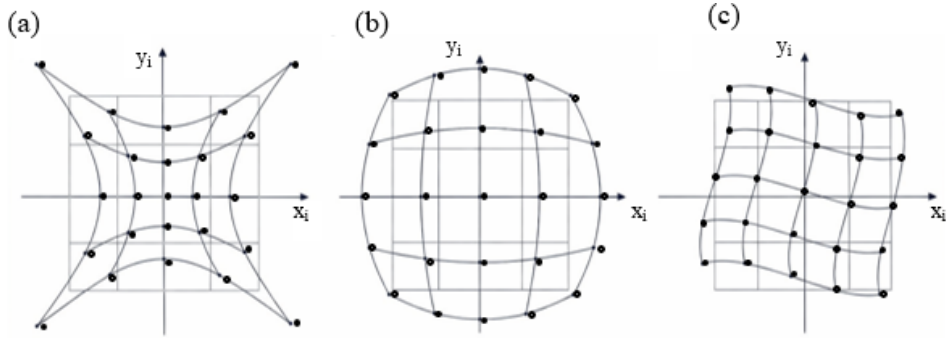

**Fig. 11.** *Image distortions. (a) Pincushion distortion, (b) barrel distortion and (c) spiral or azimuthal distortion. The mesh represents the undistorted image.*

The only dominant aberration on the optic axis is $C_3$. Though in HRTEM due to small field of view only axial aberrations are important and not the off-axial ones. However, for electron-beam lithography, off-axial aberration becomes important due to large field of view. For multipole aberration corrected lenses allowed geometrical aberrations increase significantly for a given order.

Another approach to describe aberration is based on curved wave surface and the relationship between these two provides a physical meaning of the aberrations on the image plane. This is the connection between the image aberration $W(w_0, \omega)$ and the geometrical wave front $S(w_0, \omega)$. Ideal geometrical wave surface is spherical $S^0(w_0, \omega)$ [Fig. 12]. For a non-ideal system, the spherical wave surface is warped. A small deviation from a spherical surface, the warped surface can be described by super position of ideal spherical wave surface $S^0(w_0, \omega)$ and the function $W(w_0, \omega)$ which considers the deviation. $W$ is called the aberration function, where $S(w_o, \omega) = S^0(w_o, \omega) + W(w_o, \omega)$. $W$ depends on the position of the object point and ω in the aperture plane. For Siedel aberration W can be expanded as [Eq. 3]

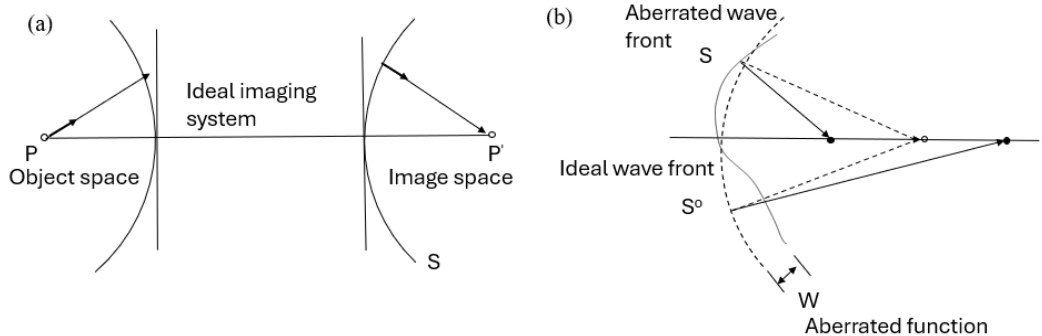

**Fig. 12.** *Wave aberration as warping of wave surface. (a) Ideal imaging system where all rays merge at point P´ from object point P. (b) Presence of aberration manifest warping of wave surface leading to non-ideal imaging system.*

$$W(w_0, \omega) = \Re \left[ \begin{array}{l} \frac{1}{2} C_1 \omega \bar{\omega} + \frac{1}{4} C_3 (\omega \bar{\omega})^2 + B_{31} \omega \bar{\omega}^2 w_0 + \frac{1}{2} F_{32} \omega \bar{\omega} w_0 \bar{w}_0 \\ + \frac{1}{2} A_{32} \bar{\omega}^2 w_0^2 + D_{33} \bar{\omega} w_0^2 \bar{w}_0 \end{array} \right] \quad (3)$$

For HRTEM imaging, small field of view off-axial aberration is negligible and after ignoring off axial aberration, $W$ reduces to

$$W(w_0, \omega) = W(0, \omega) = \chi(\omega) = \Re \left[ \frac{1}{2} C_1 \omega \bar{\omega} + \frac{1}{4} C_3 (\omega \bar{\omega})^2 \right] \quad (4)$$

Which takes the familiar form of PCTF. However, for aberration corrected microscope where $C_3$ can be made very small, say of the order of $\mu m$ compared to $mm$ in the conventional uncorrected HRTEM. Therefore, the consideration of only two axial aberrations terms is not sufficient. In aberration corrected microscope, multipole lenses allow a greater number of aberrations to be generated from symmetry. Besides mechanical imperfections, extra lenses add additional parasitic geometrical aberrations. Thus, in an aberration corrected microscopy, higher order aberrations need to be considered.

As an example, following equation shows all feasible axial aberrations up to 7[th] order is given by Eq. 5 and Table 1 list the details of the aberrations.

$$\chi(\omega) = \Re [A_0 \bar{\omega} + \frac{1}{2} C_1 \omega \bar{\omega} + \frac{1}{2} A_1 \bar{\omega}^2 + B_2 \omega^2 \bar{\omega} + \frac{1}{3} A_2 \bar{\omega}^3 + \frac{1}{4} C_3 (\omega \bar{\omega})^2 + S_3 \omega^3 \bar{\omega} +$$

$$\frac{1}{4} A_3 \bar{\omega}^4 + B_4 \omega^3 \bar{\omega}^2 + D_4 \omega^4 \bar{\omega} + \frac{1}{5} A_4 \bar{\omega}^5 + \frac{1}{6} C_5 (\omega \bar{\omega})^3 + S_5 \omega^4 \bar{\omega}^2 + R_5 \omega^5 \bar{\omega} + \frac{1}{6} A_5 \bar{\omega}^6 +$$

$$B_6 \omega^4 \bar{\omega}^3 + D_6 \omega^5 \bar{\omega}^2 + F_6 \omega^6 \bar{\omega} + \frac{1}{7} A_6 \bar{\omega}^7 + \frac{1}{8} C_7 (\omega \bar{\omega})^4 + S_7 \omega^5 \bar{\omega}^3 + R_7 \omega^6 \bar{\omega}^2 + G_7 \omega^7 \bar{\omega} +$$

$$\frac{1}{8} A_7 \bar{\omega}^8 \quad (5)$$

**Table 1:** List of axial aberrations up to 7[th] order.

| Aberration | Symbol | Wave Aberration (R) |
|---|---|---|
| Beam/Image shift | $A_o$ | $A_o \bar{\omega}$ |
| Defocus | $C_1$ | $\frac{1}{2} C_1 \omega \bar{\omega}$ |
| Twofold astigmatism | $A_1$ | $\frac{1}{2} A_1 \bar{\omega}^2$ |
| Second-order axial coma | $B_2$ | $B_2 \omega^2 \bar{\omega}$ |
| Threefold astigmatism | $A_2$ | $\frac{1}{3} A_2 \bar{\omega}^3$ |
| Third-order spherical aberration | $C_3$ | $\frac{1}{4} C_3 (\omega \bar{\omega})^2$ |
| Third-order star-aberration | $S_3$ | $S_3 \omega^3 \bar{\omega}$ |
| Fourfold astigmatism | $A_3$ | $\frac{1}{4} A_3 \bar{\omega}^4$ |
| Fourth-order axial coma | $B_4$ | $B_4 \omega^3 \bar{\omega}^2$ |
| Fourth-order three-lobe aberration | $D_4$ | $D_4 \omega^4 \bar{\omega}$ |
| Fivefold astigmatism | $A_4$ | $\frac{1}{5} A_4 \bar{\omega}^5$ |
| Fifth-order spherical aberration | $C_5$ | $\frac{1}{6} C_5 (\omega \bar{\omega})^3$ |
| Fifth-order star-aberration | $S_5$ | $S_5 \omega^4 \bar{\omega}^2$ |
| Fifth-order rosette aberration | $R_5$ | $R_5 \omega^5 \bar{\omega}$ |
| Sixfold astigmatism | $A_5$ | $\frac{1}{6} A_5 \bar{\omega}^6$ |
| Sixth-order axial coma | $B_6$ | $B_6 \omega^4 \bar{\omega}^3$ |
| Sixth-order three-lobe aberration | $D_6$ | $D_6 \omega^5 \bar{\omega}^2$ |
| Sixth-order pentacle aberration | $F_6$ | $F_6 \omega^6 \bar{\omega}$ |
| Sevenfold astigmatism | $A_6$ | $\frac{1}{7} A_6 \bar{\omega}^7$ |
| Seventh-order spherical aberration | $C_7$ | $\frac{1}{8} C_7 (\omega \bar{\omega})^4$ |
| Seventh-order star-aberration | $S_7$ | $S_7 \omega^5 \bar{\omega}^3$ |
| Seventh-order rosette aberration | $R_7$ | $R_7 \omega^6 \bar{\omega}^2$ |
| Seventh-order chaplet aberration | $G_7$ | $G_7 \omega^7 \bar{\omega}$ |
| Eightfold astigmatism | $A_7$ | $\frac{1}{8} A_7 \bar{\omega}^8$ |

Various letter symbols represent a particular type of aberration. For example, C is the spherical aberration, B is the coma, S is the star, D is the three lobes, R is the rosette, F is the pentacle, G is the chaplet aberrations. Image plot showing the modulation of wave surface depending on various types of aberration function can be found in Ref.[27].

There is another notation given by Krivanek [27]. In these notations $C(n, N)$, the first numerical subscript $n$ denotes the order $n$ of the aberration, while the second numerical subscript $N$ denotes its symmetry. In practice, whichever notation is employed, the basic physical principle remains the same. This approach is useful to describe the combination-aberrations of an aberration corrected microscope.

Now it is natural to proceed to link the relationship between the image aberration $\Delta w_i$ and the aberration function $W$ on the geometrical wave surface $W$ or alternatively by $\chi$. This is the connection between the warping of the wave surface and its impact on the image. The electron trajectories are perpendicular to the wave surface and merge to a point called Gaussian image point on the image or detector plane. If the curvature is warped, then electron trajectories do not coincide with a point and results in formation of astigmatic image. Thus, it must be the gradient of the wave surface which provides information of the ray on the image plane passing through aperture plane at an angle ω. For spherical wave surface W = 0 as well as gradient of W. Greater the gradient of $W$ larger the displacement from the Gaussian image point.

$$\Delta w_i = -M\left(\frac{2\partial W}{\partial \overline{\omega}}\right) = -M\left(\frac{\partial W}{\partial \theta_x} + i\frac{\partial W}{\partial \theta_y}\right) \tag{6}$$

Whereas delocalization as given by Eq. 7 has the same meaning.

$$\Delta r = \frac{1}{\lambda}\nabla\chi(q) = \lambda q(C_1 + C_3\lambda^2 q^2) \tag{7}$$

### D. Aberration Correctors

The performance of an aberration corrected transmission electron microscope operating at 300 keV with spatial resolution ~50-100 pm is far off from the diffraction limit of ~ 1 pm. The most dominant axial aberration of a round magnetic lens is $C_3$. In an uncorrected microscope the effect of $C_3$ can be minimized either by decreasing the focal length ($f$) or the wavelength ($\lambda$) of the beam. However, both are limited by the requirement of finite pole piece gap for the microscope column and beam damage to the sample, respectively.

Requirement of lateral resolution involves certain scattering angles to be considered for the imaging. Thus, the imaging process is non-paraxial and moreover due to imperfect lens resolution is dominated by aberrations. Aberration correction was one of the routes actively pursued to improve the spatial resolution of TEM. In 1956, Otto Scherzer laid down the conditions to overcome unavoidable spherical and chromatic aberration in round magnetic lens. These are known as Scherzer theorem [28]. Development of non-round multipole lenses proceeded with the aim to add *negative spherical aberration* and counterbalance the *positive spherical aberration* of the round objective lens [26],[27]. The first aberration corrector or Scherzer corrector consisted of two round lenses, a stigmator, two electrostatic cylindrical lenses as core unit and three octupole lenses to compensate for the $C_3$. It was built by Robert Seelinger in 1951 only to show the proof of principle that $C_3$ can be corrected through compensation by generating negative $C_3$ but not to improve the actual resolution of the microscope. The effort was limited due to severe mechanical instability associated with the instruments at that time. At 25 kV microscope, Seelinger could show 6% improvement on the spherical aberration [29]. Later in 1956, Gottfried Möllenstedt used a more stable 40 kV microscope and enlarged objective aperture to 20 mrad to demonstrate resolution improvement by a factor of 7 [30].

After a long 40 years of efforts behind the development of aberration corrected TEM, eventually hexapole corrector system for HRTEM [31–33] and quadrupole-octupole system for STEM were first implemented, the later one can correct for both the spherical and chromatic aberrations [31,34]. Though the original proposal was to develop a hexapole corrector for probe forming system and octupole corrector for broad beam system, however, their implementation appeared to be opposite. This long journey of evolution of transmission electron microscopes from the time of Ernst Ruska's construction of first TEM to modern day aberration corrected HRTEM is often compared with the milestone of Wright brother's first flight in 1903 to first manned mission to moon in 1969. Simultaneously, the improvement of objective lens was being pursued given the difficulty and challenges posed by the technological development at that time to develop a practical aberration corrected system. Present day HRTEM microscope without a corrector with $C_s = 0.5 - 2\ mm$ (objective lens) can provide reasonable HRTEM images with a spatial resolution ~ 2-3 Å.

Modern day aberration correctors are based on magnetic multipole lenses. These lenses make use of the effect of either two extended hexapole or combination of octupole quadrupole elements. Magnetic lens with multiplicity $2m$ having $2m$ poles arranged in an equidistance azimuthal angle [Fig. 13], neighboring poles have the opposite polarity. The primary effect of multipole lens is to induce aberration of order $m - 1$ with symmetry $m$. For example, for hexapole ($m = 3$) and quadrupole ($m = 2$) lenses, the primary effects are three-fold ($A_2$) and two-fold ($A_1$) astigmatisms. Therefore, they can be used to correct $A_2$ and $A_1$, respectively. However, employing these lenses as aberration correctors involves improving the resolution of the instrument, which requires a complex set up and will be discussed subsequently.

For quadrupole lens [Fig. 13 (a)], it is divergent along $x$ and convergent along and $y$, thus forming elliptical beam shape. The focusing power is stronger compared to that of round lens and already employed in particle accelerators and storage rings in synchrotron facilities. Two quadruples are required to form a round beam. The main purpose of quadrupole elements is to induce elliptical beam for octupole elements to induce negative $C_s$. On the other hand, the hexapole lens does not have any focusing action. It has been used as stigmator for three-fold astigmatism ($A_2$). Fig. 13 (b) shows the action of hexapole lens on rays away from the optic axis. Though deflection is same for rays but is function of azimuthal angle. Radial and azimuthal dependence causes the beam to end up with three-fold symmetry, hence hexapole field causes $A_2$. But combination aberration can occur for hexapole lenses stacked in a series or for an extended one where the combined resulting aberration does not belong to any individual element. Combination aberration can occur not only for hexapole, but for two optical elements of arbitrary symmetry. For two hexapole or extended hexapole the divergence between two ray paths occurs, which is the origin of negative $C_s$ [Fig. 14], which can be used for aberration correction but introduces large $A_2$. The induced negative $C_s$ is the combined aberration.

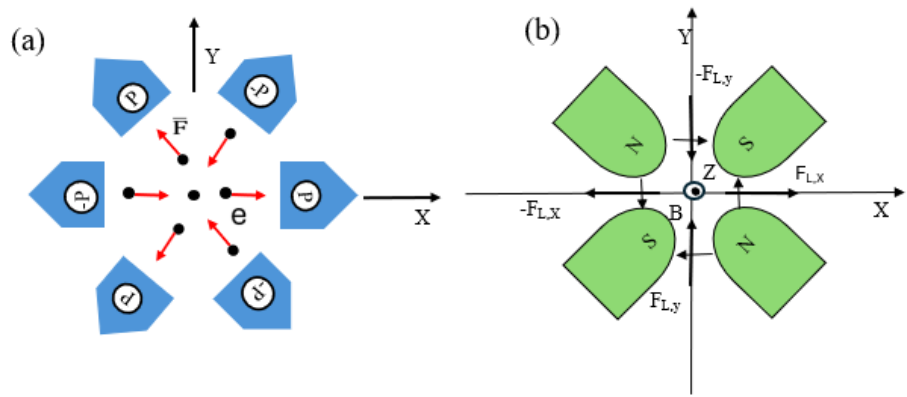

**Fig. 13.** *Example magnetic multipole lenses (a) a quadrupole and (b) a hexapole.*

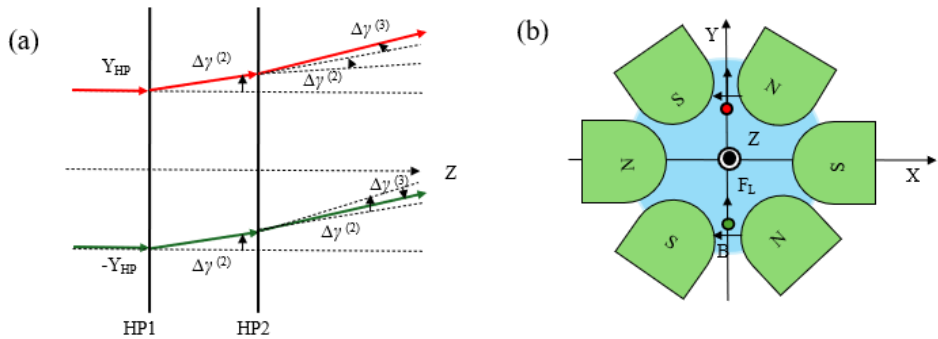

**Fig. 14.** *(a) Extended hexapole system showing electron trajectories through the long passage leading to generation of negative $C_S$. (b) cross sectional view of magnetic Lorentz forces acting on the electrons.*

The primary effect of octupole lens is four-fold astigmatism $A_3$. The symmetry of octupole lens is 4 and order of aberration is $4 - 1 = 3$. Combination aberrations of quadrupole-octupole system can be employed to correct all the third order effect i.e. spherical aberration ($C_3$), star aberration ($S_3$), and fourfold astigmatism ($A_3$). Both positive and negative $C_3$ can be generated depending on the directions. Though hexapole combinations and quadrupole-octupole combinations can be employed to generate negative $C_3$, but also create large $A_2$ and $A_3$, respectively, which need to be tackled for using them as aberration correctors. In another development the quadrupole-octupole system for HRTEM use both magnetic and electrostatic lens and corrects for both geometrical and chromatic aberrations [35,36]. The core unit of this corrector are electric-magnetic quadrupole elements. Fig. 15 shows such a unit where a magnetic quadrupole and an electric quadrupole are rotated by 45º relative to each other. The principle is similar to a Wien filter. The appropriate force balance between electric and magnetic fields on electrons corrects for chromatic aberrations along one direction only and requires two successive such units to correct for chromatic aberration along another perpendicular direction. Spherical aberration is corrected by superimposing octupole field which can accommodate astigmatic beam. Fig. 16 shows the typical location of correctors for image and probe forming systems at post and pre-specimen locations down the column, respectively.

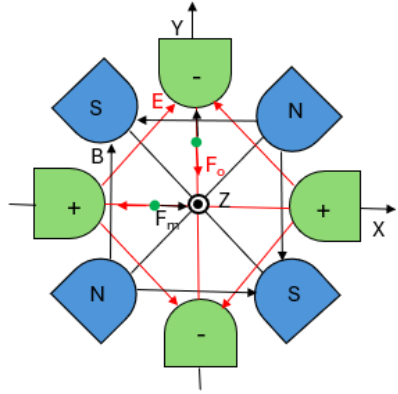

**Fig. 15.** *Schematics of an electric-magnetic quadrupole correctors for the chromatic aberration. It acts as a focusing element and Wien filter.*

Another important feature is that geometrical aberrations add up i.e. warping of the wave surface is additive. This is known as 'Additive Theorem' [37]. Therefore, there is no need to make a lens free from aberration, but corrector elements can be arranged in such a way that unwanted aberrations can be cancelled at the end e.g., positive $C_s$ can be annulled by negative $C_s$. However negative $C_s$ comes at the expense of large $A_2$ with odd azimuthal symmetry. This deteriorates the imaging performance and thus the benefit of negative $C_s$ is annulled. Hence a setup is required which only keep negative $C_s$ but other aberrations are annulled. Various designs were attempted for this. Finally Rose corrector was successfully implemented in all the early aberrated corrected instruments. Fig. 17 is the schematic arrangement of various lens elements of a Rose corrector for both image and probe forming system. It consists of two transfer doublets and two hexapole lens. The first transfer doublet projects the coma free point of objective lens into the first hexapole and the second transfer doublet inverts the magnification. Cross over through the corrector system can be thought of as inversion of magnification $M = -1$. $C_3$ is isotropic and invariant to an inversion, but $A_2$ has the odd azimuthal symmetry and is not invariant to an inversion, thus, $A_2$ will be annulled. In this setup, three lobes $D_4$ also vanished but not $A_5$ but yield spatial resolution better than 0.8 Å. To push spatial resolution down to 0.5 Å, correction of $A_5$ is necessary [38,39].

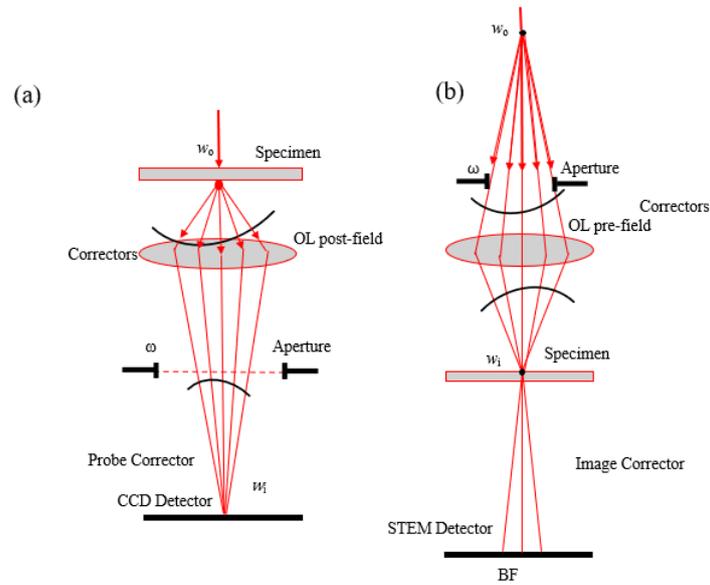

**Fig. 16.** *Location of correctors for (a) image corrector and (b) probe corrector systems.*

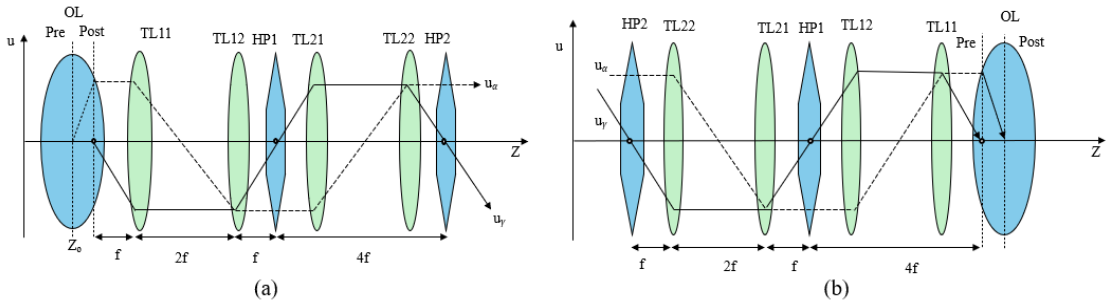

**Fig. 17.** *Rose correctors for (a) image forming and (b) probe forming systems.*

### E. Aberration corrected imaging and negative Cs phase contrast microscopy

In a conventional TEM having round lens without any aberration correctors, the value of $C_3$ is in the range of 0.5 – 1.5 mm. In an aberration corrected microscope the value of $C_3$ can be set to 0 or to some ± values and thus higher order aberrations become resolution limiting. However, the state of aberration is not permanent and need to be measured and corrected to the desired level before performing actual imaging experiments. The procedure of measurement and compensation of aberrations is known as 'aberration diagnosis'. This involved collection of a series of diffractogram of amorphous thin carbon specimen at various tilt and azimuthal angles in a specific format known as Zemlin Tableau [Fig. 18(a)] [40,41]. The method was originally developed to minimize the axial coma and later extended to the aberration corrected microscopy. Such a diffractogram is equivalent to the modulus of a two-dimensional contrast transfer function. Axial aberrations increase with the tilt and consequently change the coherent contrast transfer function. The spacings and sequence of the diffractogram depends on the applied $C_1$ and $C_3$. Ellipticity reflects the two-fold astigmatism $A_1$. Only $C_1$, $C_3$ and $A_1$ can be measured from a single diffractogram and not the impact of others. Impact of other aberrations can be obtained from the systematic tilt tableau. Fig. 18(b) pictorially shows the effect of individual aberrations on the shape of the diffractogram. For a well corrected microscope all the diffractogram appear circular for all the tilt and azimuthal angles. One can notice that even at very large angle, $C_3$ is zero after appropriate correction and only higher order aberrations are present. The procedure of correcting aberrations is as follows; at first the lower order aberrations are corrected with small tilt and then higher orders are corrected with larger tilt angle. After some trial and error and going back and forth between small and higher tilt angles the aberration coefficients can be minimized to the desired spatial resolution criteria. The procedure mentioned above is applicable for axial aberrations, not for off axial aberrations. However, off axial aberrations can be measured by taking such tableau from different regions of the object. A typical Zemlin tableau for an aberration corrected microscope before and after aberration corrections are shown in Fig. 18 (a)&(c), respectively. Various analysis for the optimum phase plate for a corrected microscope is given by Lentzen [42]. Ronchigram method is another way to diagnose and correct aberrations widely used for a probe forming system [43].

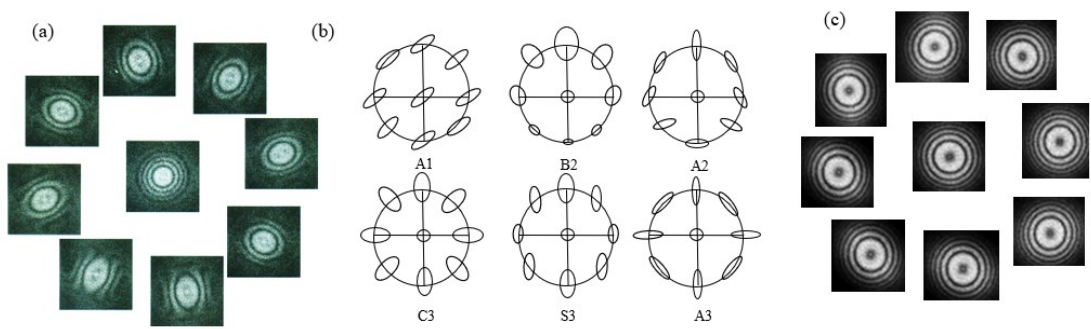

**Fig. 18.** *(a) Zemlin tableau of diffractogram from thin amorphous carbon specimen. (b) Image aberrations for various aberration types. (c) Corrected Zemlin tableau if diffractogram.*

The possibility of set the value of $C_3$ to 0 and values on either side provides an opportunity to modify the sign of PCTF function and record images of the atoms directly either with bright or dark atom contrast [Fig. 19]. White atom contrast set up gives enhanced contrast compared to dark atom contrast set up due to addition of non-linear image terms with the intensity [44]. White atom image contrast has been utilized to image directly the light atoms and distinguishing them, imaging all light and heavy atoms together, measuring ferroelectric polarization field by precisely measuring atom locations and is further discussed in sec II.B.

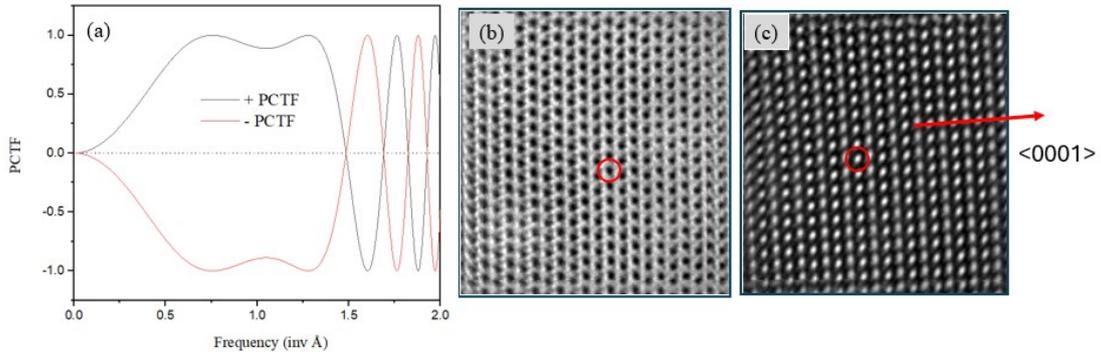

**Fig. 19.** *(a) Positive and negative PCTF corresponding to dark and white atom contrast and corresponding HRTEM images of ZnO along 11-20 directions showing the reversal of contrast depending on the sign of PCTF, Zn atom as (b) dark and (c) white dots.*

### F. A comparison with atomic resolution off-axis electron holography

Another widely used phase contrast technique in transmission electron microscope is off-axis electron holography. At medium resolution the technique is useful for electric and magnetic field mapping [21]. However, with the development of double biprism set up the technique has reached atomic resolution with sufficiently wide field of view [3,45]. The double biprism setup eliminates Fresnel fringes and the Vignetting effect essential for recording good quality holograms throughout the field of view, which is usually small at atomic resolution. There have been reports on the comparison between the techniques at atomic resolution phase reconstruction from the same sample [3,22]. One major advantage of off-axis electron holography over HRTEM/in-line holography is that the former technique transfers all the frequency information up to the information limit whereas the latter is limited by the band pass of spatial frequency as imposed by the PCTF function [Fig. 20]. However, it depends on the theoretical modelling methods to overcome any such limitations posed by the present state of the knowledge. Another advantage of off-axis electron holography is that it gives direct phase information from in and around the atoms and provides an opportunity to experimental determination of electron density. However, there are reports to obtain similar information based on HRTEM method as well [18,46].

Holography as a technique was first proposed by Gabor (1948) with the aim to get rid of the geometrical aberrations from the recorded electron micrographs. The idea was first to interfere the object wave with the reference wave and record the resulting interference pattern which is known as hologram. Such a hologram contains all the information about the object and the imaging system. During Gabor's time, the performance of electron microscope was very poor. Gabor was awarded Noble prize for the proposal (1971). However, Gabor had to abandon his method because he could not separate the direct image component from the twin image components. Subsequently, Leith and Upatnieks (1962), and Lohmann (1956) developed the off-axis geometry which could separate the twin image components from the direct part and holography technique found many wide ranges of applications in various branches of science and engineering. Practical off-axis electron holography technique was developed by Möllenstedt and Düker, who implemented an electrostatic biprism for the electron interference [3].

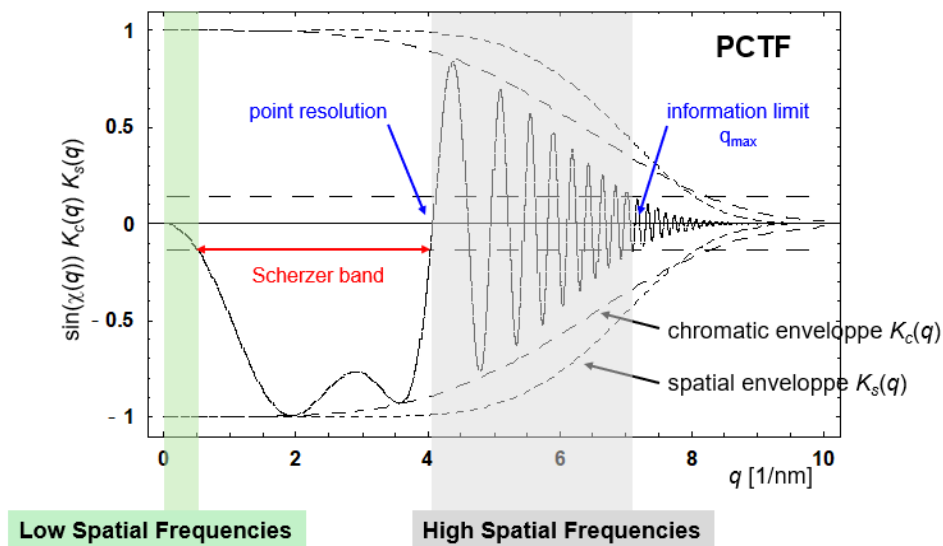

**Fig. 20.** *Comparison between HRTEM and off-axis electron holography in terms of range of frequencies the information can be transferred.*

The experimental phase retrieval methods are markedly different between the two techniques. In the atomic resolution off-axis electron holography technique, electron interference fringes encode phase information at a subatomic length scale. In standard practice, Fourier transformation of the off-axis electron hologram results in three bands, one central (CB) and two side bands (SBs), where the side bands are complex conjugates of each other. Phase information can be directly extracted from one of the two SBs. The CB is equivalent to in-line holography, which contains mixed amplitude and phase signals. On the other hand, the retrieval of phase shift from in-line holography requires a series of images to be recorded at different focus values. Various reconstruction schemes have been developed to retrieve the object exit wave (OEW) function from the experimental through focal image series method [1]. The development of both the experimental approaches to extract phase information dates back to the BRITE EURAM program [47]. Comparisons of phase information by two different methods have been reported by a few groups both at medium and atomic resolutions. However, quantitative phase information obtained so far through off-axis and in-line holography does not correspond to each other for the same sample area and depends on the frequency range considered for the analysis [22].

Quantitative imaging is a recent area of active research in the atomic resolution electron microscopy community, and therefore, understanding the accuracy of the experimental phase determination and its correlation with the property of materials is crucial for its success and contribution to the materials and microscopy sciences. Both aberration-corrected HRTEM and atomic resolution off-axis holography provide a unique opportunity to study phase information at the atomic and subatomic length scale [2].

## II. Applications in material science
### A. Direct imaging of light atoms and qualitative interpretation

In a $C_s$ corrected microscope the sign of $C_s$ can be set to zero or either to a positive or negative value. Thus, the sign of PCTF can be reversed accordingly and direct imaging of atom column can be set up with bright or dark atom contrast. Fig. 19 showing the dark and white Zn atom contrast in ZnO recorded in JNCASR aberration corrected microscope. The corresponding values of $C_s$ and $\Delta f$ are indicated in the figure. The concept of negative $C_s$ HRTEM imaging was first demonstrated by K. Urban [44]. The direct imaging of all the atomic columns in SrTiO$_3$ crystal together in single micrograph can be recorded by this imaging technique. In another powerful example is given in YBa$_2$Cu$_3$O$_7$ superconducting crystal by direct imaging of all the heavy and light atoms together. This example shows clearly the O vacancy ordering in the crystal by imaging the crystal along two orthogonal directions and revealing the presence and absence of O atomic columns [16]. In another similar approach, Iijima et al imaged all B, N atoms in BN monolayer with vacancy defects and edge structures [17]. It was possible to identify the individual B and N atoms directly from the HRTEM image intensity contrast directly which is akin to the chemical information generally obtained from simultaneous spectroscopic signal.

Among various other example of direct interpretation of atomic length scale material structure are given in MoS$_2$ thin film grown on sapphire substrate showing interlayer stacking and interfacial atomic bonding, cation ordering in inverse spinel NiFe$_2$O$_4$, CoFe$_2$O$_4$ showing co-existing regular inverse spinel structure and A site cation defects, MoS$_2$-ReS$_2$ 2D alloy showing 2H to 1T$_d$ phase transitions etc, can be found in Ref. [10,13]. However, it will be necessary to perform conventional through focal series acquisition or applying direct method to count number of atoms in each column and identifying various atomic species. There are reports on measuring electrostatic potential from atomic length scale [46].

### B. Atom position and ferroelectric polarization

The power of direct atom imaging with bright atom contrast by NCSI imaging condition was vividly demonstrated for various ferroelectric materials [48,49]. In a ferroelectric material the relative internal shift of atoms gives rise to electric polarization field which are useful for memory applications like magnetic domains in a ferromagnetic material. Under NCSI imaging condition detecting O atom shift the magnitude of polarization, domains and domain boundaries are determined directly. Following the method Na0.5Bi0.5TiO3 system was investigated to find out whether the thermodynamically allowed structure is rhombohedral or monoclinic [50,51]. Monoclinic distortions were found to be associated with local octahedral tilts in phase which removed after poling and go to rhombohedral structure. In another investigation metastable monoclinic and orthorhombic phases and electric field induced irreversible phase transformation in BaTiO3 was studied. Displacement of the Ti [001] direction corresponds to the tetragonal phase and the [011] direction corresponds to the orthorhombic phase.

## C. Quantitative imaging: atom counting, potentials, orbitals

As already mentioned, HRTEM imaging method is routinely used for imaging lattice planes, crystalline quality, relative lattice parameter measurement, atomic resolution imaging of various crystallographic defects etc. These details are directly interpretable and reliable if in conjunction with NCSI imaging conditions are used. However, effort is ongoing to fully realize the potential of atomic resolution phase contrast technique beyond intuitive structural interpretation e.g., chemical identification of atom, atomic resolution tomography of crystalline materials, single dopant chemical bonding or charge state etc. We elaborate some of the notable examples based on different atomic resolution phase contrast techniques from the literature. Here are the few examples of quantitative imaging by traditional HRTEM imaging method. Ute kaiser et al first analyzed the charge distribution due to N doping in graphene by aberration corrected HRTEM [18]. Single e- charge transfer is shown from N to surrounding C atom in N doped graphene by HRTEM and combined DFT calculation. In another report, 3D arrangement of all the atoms, dopant & defects & atoms on the surface of MgO nanocrystal have been reconstructed from a single 2D HRTEM image. The method is based on applying a dedicated numerical data analysis method [15]. Big bang tomography is the new theoretical model proposed by Van Dyck et al. where all C positions in bilayer graphene are reconstructed from the single HRTEM image. In this case the phase velocity analogous to expanding universe concept has been used [52]. These are only few examples & the area is still open to newer & simpler methods.

However, based on our own research experience I am highly lenient to HRTEM based technique despite drawbacks like in-line type holography method, highly sensitive image patterns with experimental parameters like defocus, thickness of specimen etc. In our experience and opinion HRTEM based phase contrast techniques may be more powerful in extracting many useful information depending on the methods to be employed to extract information. Another advantage is that it may be possible to extract information from a single image recorded under moderate exposure time which is beneficial for radiation sensitive materials.

## D. Image simulation

Both image simulation and reconstruction procedures are integral parts of quantitative phase contrast imaging based on HRTEM and complementary techniques. The entire topic of quantitative HRTEM falls into two broad categories; (*i*) object exit wave (OEW) reconstruction to retrieve the missing phase information from the recorded image intensity pattern, and (*ii*) image simulation to interpret the OEW with the object structure. Probe electron waves after propagating through the specimen thickness undergoes modulation into its phase and amplitude and after exiting the specimen is known as OEW [Fig. 21]. OEW encodes information on the specimen. The OEW can be recorded as intensity pattern either in the image plane or in the diffraction plane. The phase information encoded by OEW cannot be read directly from such intensity patterns and involves image reconstruction analysis. Various reconstruction procedures involving HRTEM and off-axis electron holography are discussed briefly in the next section. Two types of phases are of importance here, (*i*) crystallograp1hic phase which gives information on the arrangement of scatterers, and (*ii*) object phase which gives information on the strength of the scatterers. The object phase helps in identifying type and number of atoms in an atomic column. For detailed discussion on the simulation and reconstruction methods and our contribution in the field reader is requested to refer to Ref. [1,2] and supplementary therein.

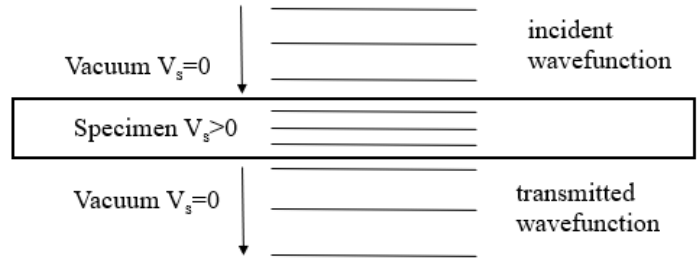

**Fig. 21.** *Transmitted wave function after probe propagated through the specimen thickness is called object exit wave (OEW) function.*

Atoms are generally treated as electrostatic potential object and the corresponding electric phase shift of probe electron wave and its measurement at atomic and medium resolution is the basis of the phase contrast electron microscopy techniques. However, the change in phase ($\phi$) of the probe electron wave after interaction with the specimen potential and resulting modulation in intensity pattern has been treated in fundamentally distinctive ways, e.g., (*i*) transmission function based on weak phase object approximation (WPOA) along with Zernike type $\pi/2$ or $\lambda/4$ phase plate equivalent to phase contrast transfer function (PCTF) to account for the lens aberration, where the phase change is incorporated in terms of change in magnitude of momentum vector $k$ of the probe electron due to specimen potential, (*ii*) phase change according to scattering amplitude in terms of atom scattering $f(k)$ and structure factor $F(g)$ along with PCTF, and (*iii*) self-interference in HRTEM and holographic fringe shift in off-axis electron holography where a phase term (± φ) is added inside the trigonometric functions with respect to the reference phase. Kindly note that the phase change due to aberration through PCTF is not added with the object wave rather it is applied as a frequency filter and point spread function (*psf*) in the diffraction and image planes, respectively. Our study indicated that different ways of treating atoms lead to different results for the same atom type and this was the primary motivation for us to propose alternative methods of image simulation and reconstruction.

First, a brief discussion is provided on the existing methodologies involved in the image simulation of atoms. Subsequently, an alternative atom image simulation method is introduced where the geometry of interference based on the direction of momentum vector is emphasized. The method is based on atomic potential center as an interferometer akin to the electron biprism within a short range of focus variation (<10 nm) from the reference Gaussian image plane and resembles Abbe's picture of diffraction. In this alternative method, the aspect of phase change has been treated like off-axis electron holography considering the wave interference at an angle, and its analogy with other approaches can be understood with the help of interference geometry and associated momentum vector direction. Simulation results are compared with the experimental image of 2D materials of MoS$_2$ and BN recorded under specific settings of third order spherical aberration $C_s = -35$ μm and defocus $\Delta f = 1, 4,$ and 8 nm and are found to be in good agreement.

In TEM, the interaction of probe electrons and specimen and subsequent flow of information is based on the principle of wave optics. In HRTEM the electron illumination is considered as a plane wave while interacting with the specimen. A description will now be provided on how the phase information is encoded and flows between the image and diffraction planes both in standard optics and electron diffraction in atomic system following few available concepts.

The interference geometry of Fresnel and Fraunhofer diffraction in standard optics describes the scattering distributions in image and diffraction planes, respectively. The intensity pattern can be calculated for both the regimes by considering Fresnel-Huygen's construction. This is based on the consideration that every point of the wavefront is the source of secondary spherical wavelets & wavefront at some propagation distance is given by the envelop of the wavelets. The geometry of diffraction and the corresponding path difference is distinct between the Fresnel and Fraunhofer regimes. In Fresnel's diffraction geometry, the path difference is obtained between wave vectors with respect to the outer curvature of the sphere being converged to an imaging point. However, the interference geometry for Fraunhofer is the path difference between the wave vectors from the various spatial points at the aperture opening having same momentum direction towards the diffraction plane. Similarity can be noticed between the rectangular slit diffraction in light optics and Fraunhofer electron diffraction pattern in TEM.

However, the image of an object/ aperture (not Fresnel) at the far field can also be calculated by FT of the aperture function $f(x,y)$. Abs (FT) of the $f(x,y)$ is diffraction pattern. This is also the foundation of Abbe's theory of imaging where the point of interaction between the object and probe gives off Fourier waves with a range of frequencies. Higher the frequency, higher will be the scattering angle from the optic axis. This is also the basis of resolution criteria, where restricting the highest possible frequencies allowed by numerical aperture limits the resolution of the imaging system. Another important point is that the diffraction geometry in Abbe's theory has similarity to off-axis holography if we consider only a pair of frequencies across the mirror plane and has close similarity with the alternative image simulation method introduced by us. In the case of atomic systems, the electrostatic potential replaces the slit object function. The FT of isolated and periodic atomic potential in the diffraction plane is the atom scattering factor $f(k)$ and structure factor $F(g)$, respectively. Now we describe image of the atoms as calculated by different approaches.

In image simulation based on the Zernike phase contrast theory, the object transmission function for pure phase or weakly scattering object is represented by a complex function of the form $F(x) = e^{i\phi(x)}$ or $\sim 1 + i\phi(x)$ for small $\phi$, where $\phi$ is a real phase function corresponding to the discrete or periodic transparent object. This is known as weak phase object approximation (WPOA). In fact, the object function is real, it is the replica of the object in the form of object wave that carries the information encoded into its phase and amplitude and is similar in function to the Fourier or Abbe waves and holographic direct and twin image wave components. The effect of Zernike phase plate modifies the intensity of the object wave that depends linearly on the object phase according to $I(x) = 1 \pm 2\phi(x)$.

WPOA is a straightforward and widely applied approach to simulate the HRTEM images of thin samples. In HRTEM, WPOA describes the phase shift of probe electron wave due to object electrostatic potential projected along the beam propagation direction, and the transmission function has the following expression after invoking WPOA, i.e., ignoring the terms with $\sigma^2$ and higher order,

$$t(x) = 1 - i\sigma V_t(x,y) \qquad (8)$$

Where, $V_t(x,y)$ is the projected specimen potential and $\sigma = \frac{2\pi m e \lambda}{h^2}$ is the interaction constant. Approximation in Eq. 1 is essential to retain the object information in the image plane.

The image intensity after considering the lens effect is given by

$$I(x,y) = \psi_i(x,y)\psi_i^*(x,y) \approx 1 + 2\sigma\phi_p(-x,y) * \mathcal{F}\{sin\chi(u,v)P(u,v)\} \qquad (9)$$

Image intensity calculated by using Eq. 9 with and without considering the lens response for isolated Mo, S, N, and B atoms have been performed. Not considering lens response is similar to Zernike like phase transfer and as the potential function is asymptotic, peak intensity value will remain undefined with a background value of 1. Considering aberration through optimum PCTF ($C_S = -35 \ \mu m$ and $\Delta f = 8 \ nm$), peak values (and FWHM) of ~ 22000 (0.25 Å) and 3500 (0.25 Å) are obtained for Mo and B atoms, respectively. This gives a Mo/B peak intensity ratio of ~ 6.2. Peak intensity increases linearly with atomic number. The trend is in contrast with the experimental observation where changes in peak intensity are observed in the first decimal place with the atomic number. The high peak value according to Eq. 3 is due to the convolution procedure and cannot be normalized individually as the image without PCTF is not known.

Another approach of image simulation incorporates atom scattering factor directly instead of specimen potential. The transmitted wave function in this case can be derived from the Schrödinger integral equation and has the following form

$$\psi_t(x) = \exp(2\pi i k_z z) + f_e(q)\frac{\exp(2\pi i q.r)}{r} \qquad (10)$$

Where, $q = k - k_0$, and $f_e(q)$ is the atom scattering factor and is defined by,

$$f(q) = -\frac{m}{2\pi\hbar^2} \int V(r')e^{-2\pi i q.r} \, d^3r \qquad (11)$$

Which is the FT of the scattering potential. The solution of wave function based on the differential form of Schrödinger equation has the form of a plane wave.

This result is used along with the scattering factor as derived by Moliere to calculate the image of isolated atoms by using the following equation.

$$g(x) = \left| 1 + 2\pi i \int_0^{k_{max}} f_e(k) \exp[-i\chi(k)] J_0(2\pi kr) k \, dk \right|^2 \qquad (12)$$

Where, $f_e(k)$ is the electron scattering factor in the Moliere approximation which has the advantage over Born scattering factor due to the presence of imaginary component. $\chi(k)$ is the aberration function, $k_{max} = \alpha_{max}/\lambda$ (rad Å$^{-1}$) is the maximum spatial frequency allowed by the objective aperture and $J_0(x)$ is the Bessel function of order zero.

The phase contrast image calculated using the above expression varies weakly with atomic number and the peak phase shift $\varphi_{max}(rad)$ follows $\sim Z^{0.6} - Z^{0.7}$, where $Z$ is the atomic number. Though the trend can be complicated depending on the valence electron filling, and for specific atoms with higher $Z$ can have smaller contrast compared to atoms with lower Z next to each other in the periodic table. The peak intensity is almost the same irrespective of the atomic number and changes only in the second decimal place. The difference in peak intensity and FWHM maximum calculated by two different methods are markedly different. The intensity values calculated based on Eq. 12 show frivolous dependence considering only PCTF irrespective of atom number. However, the peak values are much smaller, and FWHM are higher by a factor of two, respectively calculated by Eq. 12 and Eq. 9, and the difference will remain even after the energy balance.

Now our proposal on the alternative image simulation method based on the concept of the atom as an electrostatic charge center and its action as an interferometer is described.

The intensity pattern of the hologram is given by Eq. 13. The details on the off-axis electron holography methods and practices can be found in Ref. [21]. The intensity pattern of the hologram is given by equation 7

$$I(x,y) = I(x) = a_1^2 + a_2^2 + 2a_1 a_2 \cos(2\pi q_c x + \Delta\phi) \tag{13}$$

Where, $q_c = 2k_x$ is carrier spatial frequency of the hologram. $a_1$ and $a_2$ are the amplitudes of waves undergoing interference at an inclination angle due to action of biprism. $x$ is the spatial coordinate variable of the interference pattern and $\Delta\phi$ is the difference in phase between the two interfering partial waves, typically acquired by one of the two waves due to object potential. $\Delta\phi = 0$ for the vacuum wave. Information on $\Delta\phi$ appears as a small deviation in a straight hologram fringe pattern.

Now, the image formation by the interference of waves due to atom charge center similar to 1D electrostatic bi prism is described here. Considering the radially symmetric atomic potential the intensity pattern can be calculated following Eq. 13 after incorporating wave interference effect from a given radial zone of extent $\Delta r$. The zones described here are similar to binary type Fresnel zone plates with multiple foci, which depends on the scattering angle with different order of magnitude. Calculating the interference pattern along all azimuthal inclination angles for the peripheral zone requires two additional considerations compared to the unidirectional interference pattern. The first consideration is that the wave flux will depend linearly on the perimeter $2\pi r$ which is a function of radial distance $r$ from the center of the atom. Larger the perimeter or the zone area [$\pi(r_2^2 - r_1^2)$] away from the atom center, higher will be the flux of the wave approaching for the interference. The relative intensity contribution at the center of the pattern from different rims belongs to the same spatial coherent zone is scaled with $2\pi r$, where $r$ is the radial distance from the center of the atom.

$$I_{rad\ int}(r) = a_1^2 + a_2^2 + 2a_1 a_2 \cos(2\pi q_c r) * 2\pi(r_{max} - r) \tag{14}$$

The second consideration is that of flux balancing between the flux of wave at the plane of the atom as given by the coherent rim area $\pi(r_2^2 - r_1^2) * I$, where $I$ is the intensity at a given pixel point within this zone, and resulting interference field over a circular area around the optic axis as given by $\pi r^2 \times I_{rad\ int}$ (1$^{st}$ law of thermodynamics).

$$\int_{-drr}^{drr} I(r)\, dr = \pi(r_2^2 - r_1^2) \times 1 \tag{15}$$

Where, $drr = (r_2 - r_1)/2$ and $r = \sqrt{x^2 + y^2}$. $I = 1$ is the minimum intensity count on a pixel of size of 1 pm considered for the present calculation.

The peak intensity has a dependence of ~ $AZ^B$, where $A$ is a fitting constant and $Z$ is the atomic number. Exponent B changes from 0.5 to 0.26 for change in focal length from 1 to 8 nm. However, the peak intensity values decrease significantly after considering image aberration.

The method puts forth a physical picture based on the geometry of interference within classical wave optics. The method can predict the absolute intensity of atoms with atomic numbers in the correct order unlike the other two methods where relative intensity between atoms can be compared. The simulated results corresponding to the image intensity are in close agreement with the experimental images of Mo and B atoms recorded under the optimum combination of third order spherical aberration $C_s = -35$ μm and defocus $\Delta f = 1$, 4, and 8 nm.

### E. Image reconstruction

As already mentioned, that the HRTEM image reconstruction techniques concern with the retrieving missing phase information from the recorded image intensity patterns. In the present review an alternative reconstruction method is presented for retrieving the object exit wave (OEW) function directly from the recorded image intensity pattern. The method is based on applying a modified intensity equation representing the HRTEM image. For a detailed comparative discussion between the existing methodologies involved in the reconstruction of OEW, off-axis electron holography and the present proposal, the readers are requested to visit Ref. [1]. Phase shift extracted from the experimental images of $MoS_2$, BN and ZnO are found to be in excellent agreement for most of the atom types investigated with theoretical reference values. A second method based on Fourier series expansion of the diffraction pattern is shown to be effective in retrieving the isolated and periodic image functions of certain forms directly. However, for aperiodic object information e.g., defects, dopants, edges etc., the first method is the suitable one. Among various methods, complexities involved in OEW reconstruction based on conventional through focal HRTEM image series have been addressed and its analogy and differences with respect to off-axis electron holography are highlighted.

In case of off-axis electron holography, the retrieval of OEW is performed first by Fourier transformation (FT) of the image containing electron interference pattern, then selecting one of the two side bands (SBs) which are complex conjugate (or twin image) to each other followed by inverse-FT. This procedure isolates the CB and the twin image wave functions from the recorded image. The phase and amplitude can be evaluated either by the arctan function corresponding to inverse-FT or fitting the inverse-FT pattern with the help of image simulation. As the starting data is the image, therefore, the FT procedure does not lead to loss of any information in terms of crystallographic phase and inverse FT can return the image intensity pattern. Deconvolution of coherent aberration envelope can be performed posteriori that modifies the aberration figure in the image plane.

However, in case of HRTEM, wavefunction and its twin image cannot be separated out like off-axis e- holography. They are superimposed on each other both in the frequency & image plane and has a non-linear term $\psi\psi^*$. Therefore, almost all the reconstruction methods in HRTEM involve multiplying the image intensity pattern recorded at different focus settings with a complex filter function consisting of a coherent aberration envelope corresponding to each focus and then summing up over all the images to eliminate the unwanted twin image and non-linear components from the wanted OEW function. The complexities involved with reducing the unwanted components from the wanted OEW function led to the development of several reconstruction algorithms.

Now, within the coherent detection i.e., multiplying by corresponding phase conjugate and summing over N images after assuming $k$ is non-zero which allows to omit the delta function and for a constant $k$, the image intensity becomes

$$\sum_n I_n \exp(-\pi i \lambda z_n k^2) = \psi \sum_n 1 + \psi^* \sum_n \exp(-2\pi i \lambda z_n k^2) +$$
$$\sum_n H_n \exp(-\pi i \lambda z_n k^2) \tag{16}$$

From the above expression, it was concluded that the wave functions $\psi$ and $\psi^*$ accumulate to N and $\sqrt{N}$ times to its original value, respectively [53]. The non-linear term $H_n$ behave randomly with the defocus. Thus, by dividing the above equation by N, the wanted OEW function may be recovered. Equation 7 is widely considered in almost all the through focus image series OEW reconstruction. And all the efforts on developing image reconstruction codes primarily deal with eliminating the non-linear and complex conjugate terms and finding the best fit with the model calculation. However, the presence of complex conjugate and non-linear terms will always be present depending on the extent their weights are subdued.

However, it is demonstrated that by marginally modifying the form of the intensity equation [Eq. 17 & 18] describing the HRTEM image and concomitant justification, it is possible to retrieve the phase information directly from the atomic resolution images. The same equation in an intermediate form based on the wavefunction formalism is used in the existing OEW reconstruction procedures in the case of in-line and off-axis electron holography. Nonetheless, there exists an alternative way of analysis, after complete evaluation of Eq. 16 corresponding to one single image, yields the following final form

$$I_{in\ line} = |\psi_0 + \psi_i|^2 = A_0^2 + A_i^2 + 2A_0 A_i \cos(\phi_i - \phi_0) \tag{17}$$

Where, $\psi$ is replaced with the wave function of the form $A(x,y)e^{i\phi(x,y)}$, describing the image intensity pattern based on self-interference between reference incident and scattered waves within the picture of single electron wave interference phenomena. The expression has the similarity with the wave interference between the reference and object waves in off-axis electron holography. However, in off-axis geometry, there is an additional phase term $Qx$ due to wave interference at an angle that gives rise to spatial modulation in the interference field [54]. Moreover, Eq. 16 is in an intermediate state, which is used to eliminate the effect of twin image and non-linear terms by working in the diffraction plane. However, the existence of the final form of the expression implies that fitting the intensity equation alone and evaluating the phase term should, in principle, allow extracting the relative phase change from the image plane. More details on associated twin image wave functions and applicability of Eq. 17 can be found in Ref. 1. The results based on the above schemes can be found in Ref. [3,55] and do not show any systematic trends with the sample thickness.

The object phase can be recovered by applying Eq. 18 as given below, which is a modified form of Eq. 17 describing the image intensity pattern in HRTEM within few nanometers from the Gaussian image plane.

$$I_{in\ line}(x,y) = |\psi_0 + \psi_i|^2 = A_0^2 + A_i^2 + 2A_0 A_i\ sin\{\phi_i(x,y)\}$$

$$= \alpha I_0 + \beta I_0 + 2\sqrt{\alpha I_0 \times \beta I_0}\ sin\{\phi_i(x,y)\} \tag{18}$$

Where, $I_0$ is the mean vacuum intensity. The factors $\alpha$ and $\beta$ represent the fractions of direct and scattered part of the intensity and can be determined by analyzing the image pattern, where $\alpha + \beta = 1$. Typical values of $\alpha$ and $\beta$ are found to be ~ 0.88/0.95 and 0.12/0.05, respectively from the experimental images of MoS$_2$/BN. The method described here works on atomic resolution HRTEM image recorded under suitable imaging conditions i.e., with a particular combination of spherical aberration coefficient $C_s$ and defocus $\Delta f$ that sets the optimum contrast and resolution.

Here is a brief comparison & justification between the formalism based on wave function & the final state of the intensity equation working with the wavefunction formalism means retrieving the probability amp from the exp. Whereas, working with the final state means working with the probability density. Probability density is the realistic approach which is also in line with the Born rule in QM. That is an essential component of Copenhagen interpretation of QM. As we all know, Born rule is an important link between the complex no based QM wave functions & it's exp measurement. Now the emphasis of working with the WF based formalism is due to attempting to separate the twin image. As we already saw in case off-axis twin images can be separated in frequency space. However, we argue that isn't an issue if one wants to work with the final form of the Eq. 18.

Now, the recovery of image function containing information on both crystallographic and object phase is described if the information is available only in the diffraction plane. The real image function $f(x)$ of both symmetric non-periodic and periodic in particular forms e.g., Gaussian, reciprocal etc. can be retrieved by cosine Fourier series expansion of the absolute FFT of such functions followed by summation over all frequencies. As already mentioned, the absolute FFT and Fraunhofer pattern are equivalent to each other. The procedure is similar to the zero-phase reconstruction using the Patterson function in X-ray crystallography applicable to smaller size molecules [56,57].

The unique feature underlying the workings of this reconstruction method is that for most functions the integration of imaginary part over spatial parameter or abs-Im component is negligible compared to real part or abs-Re. Therefore, the original function $f(x)$ can be retrieved completely by the following real cosine Fourier series alone. If $F(k)$ is the absolute FFT of $f(x)$, then $f(x)$ can be retrieved from the $F(k)$ directly via following real Fourier series expansion

$$f(x) = absolute\left(\sum_{n=-k}^{n=k} \frac{1}{n} C_k \cos(2\pi k x)\right) \qquad (19)$$

Where, $n$ = number of data points in the frequency axis, $k$ is the frequency, $C_k$ = absolute of Fourier transformations or diffracted intensity at some frequency $k$, and $x$ is the 1D spatial coordinate over which real image function will be defined.

For periodic function in 1D, $f(x) = f(x \pm px)$, where $p$ =0, 1, 2, …, one need to ensure that the range of $n$ should be $= \pm \frac{2p}{M}$, where $M$ = number of data points both in real and diffraction space. In the case of a 2D isolated function, Eq. 19 will have another summation over $n_y$ for the second orthogonal axis and the modified Fourier series equation for isolated and periodic 2D function can be found from Ref[1], where the functional form of cosine functions changes depending on the isolated 2D and periodic 2D functions in order to get an exact fit.

### F. Other phase contrast methods: DPC, iDPC

Differential phase contrast microscopy is a technique which is based on the STEM set up and was first proposed by Rose in 1974 [58]. Various variants of the techniques are available e.g., DPC, dDPC, iDPC based on the principle. Differential phase contrast (DPC) is based on measuring the difference of signals coming from two opposite segments of a divided detector. Subsequently it was pointed out that momentum transfer to electron probe due to specimen can be measured using first moment detector which yields the centre of mass (COM) of the electron beam illumination. The momentum transfer is linearly related to the gradient of the phase of the specimen transmission function. In 1979, Wendell and Chapman, established the link between COM and DPC. COM image is the cross correlation (not the convolution) between probe intensity and gradient of the specimen phase transmission function. Differentiation of COM, dCOM or dDPC are the Charge density imaging technique.

Differential phase contrast microscopy at atomic resolution has been reported [23–25,59]. The image contrast reflects the gradient of the electrostatic potential of the atoms; that is, the atomic electric field, which is found to be sensitive to the crystal ionicity. Both the mesoscopic polarization fields within each domain and the atomic-scale electric fields induced by the individual electric dipoles within each unit cell can be sensitively detected in ferroelectric $BaTiO_3$. DPC imaging, where images are formed by subtracting intensities on detector segments diametrically opposed to the optical axis. This converts the beam deflection, which relates closely to the gradient of the phase change through the specimen, into image contrast. In another example, electric field imaging of single atoms in $SrTiO_3$ crystal has been reported [24]. Real space visualization of intrinsic magnetic field in an antiferromagnet (α-$Fe_2O_3$) at atomic resolution is performed in DPC-STEM under magnetic field free condition [25]. In this special DPC-STEM microscope the residual magnetic fields generated near the sample center are <0.2 mT, some 10,000 times smaller than those in the conventional magnetic objective lenses. Another technique, Ptychography is a computational method of microscopic imaging. It generates images by processing many coherent interference patterns that have been scattered from an object of interest. The technique has been used to demonstrate super resolution of 0.39 Å in twisted bi-layer $MoS_2$ by resolving the two Mo atom separation due to misregistration between the layers [60]. With the availability of several phase contrast methods at atomic resolution opens the door for further interesting research comparing various techniques for the same materials and length scale.

### III.  Future perspective of HRTEM imaging

With the unprecedented spatial resolution achievable with the aberration corrected transmission electron microscopy, it has become imperative to ask questions on what will be the limiting applications of such an advanced technique? We already have seen direct interpretation of atomic columns, counting and identification of various atoms directly related to the atomic organizations of a material. However, can such a technique be extended to answers various nano questions or can we answer material properties by imaging? To answer material properties by imaging one must record the potential in and around the atoms by such a phase shift technique. Once the potential information is obtained one can take help from density functional theory-based approach or equivalent methods to make use of atomic orbital picture to be fitted into that to predict properties. This will require fine details of the potential to be measured and demands employing highly sensitive electron detectors and robust numerical and theoretical methods to be employed.

#### A.  Advanced electron detectors

Recording devices in TEM has undergone significant evolution both in imaging and EELS spectrometry. Introduction on electron direct detection cameras was meant to be in the field of biological imaging where ultra-low dose and at the same time sufficient image contrast is required to prevent specimen damage. Direct electron detection cameras offers both high spatial and time resolution [61,62]. Fig. 22 showing pixel density comparison between GIF CCD and K2 direct detection camera. Pixel size can be as small as 2 pm and even below. With this unprecedented high signal to noise ratio and high pixel density in data set can provide the opportunities to extract fine details about the materials.

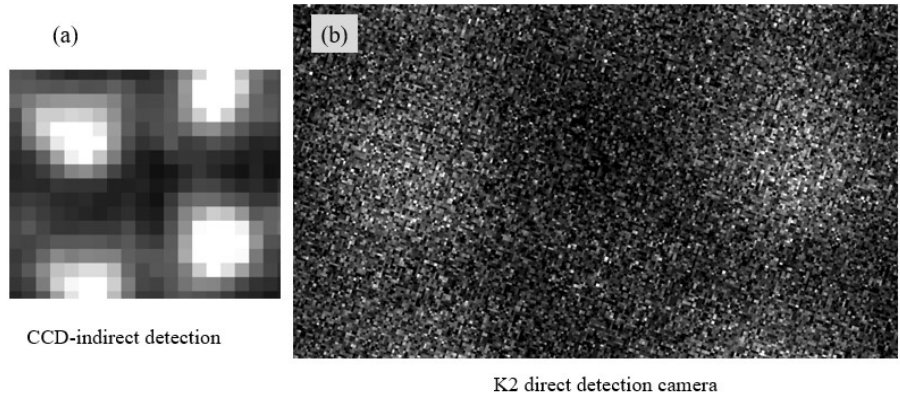

**Fig. 22.** *Pixel density in (a) Gatan Tridiem and (b) K2 direct electron detection camera. Pixel size in the image (b) is ~ 2 pm.*

## IV. Concluding remarks

The electron microscopy research group at JNCASR worked with double Cs corrected microscope equipped with a gun monochromator. The instrument had spatial resolution better than 0.8 Å and an energy resolution 180 meV operating at 300 keV. This was a great instrument to study materials at atomic and sub-atomic length scale by imaging and spectroscopy. Some examples of direct imaging are given on identifying the atom type, vacancy defects, chemical bonding state suitable for mono-layer materials. However, for materials having finite thickness, and nanocrystals having 3D shape extraction of such information involves tedious image analysis and structural modelling task. Development of different methods have been mentioned, and efforts are being made to simplify the task to extract information reliably and effectively. The spatial resolution has been achieved below 0.5 Å and with the availability of direct detection camera signal can be recorded at single electron detection level. This provides an unprecedented opportunity to study materials not only at high spatial resolution but with extremely superior signal to noise ratio. We will have to watch to see many exciting developments in analysis methods to extract useful information about the materials.